\documentstyle[11pt,amsmath,amssymb,amsfonts,theorem,float,color]{article}
\newtheorem{theorem}{Theorem}[section]

\newtheorem{cor}[theorem]{Corollary}
\newtheorem{lemma}[theorem]{Lemma}

\newtheorem{remark}[theorem]{Remark}

\newtheorem{example}[theorem]{Example}
\newtheorem{problem}[theorem]{Problem}
\newtheorem{con}[theorem]{Conjecture}

\def\qed{\hfil {\vrule height5pt width2pt depth2pt}}
\def\bref#1{(\ref{#1})}

\def\R{{\mathcal{R}}}
\def\N{{\mathbb{N}}}
\def\P{{\mathbb{P}}}

\def\startpage{1}  

\def\dim{\hbox{dim}}
\def\deg{\hbox{deg}}

\def\span{\hbox{\rm{span}}}

\def\R{{\mathcal{R}}}

\begin{document}
\renewcommand{\baselinestretch}{1.2}
\setcounter{table}{0}

\title{On Functional Decomposition of\\ Multivariate Polynomials with \\ Differentiation and Homogenization$^1$}
\author{Shangwei Zhao, Ruyong Feng and Xiao-Shan Gao \\
        KLMM, Academy of Mathematics and Systems Science\\
        Chinese Academy of Sciences, Beijing 100190, China  }
\date{ }
\maketitle

\footnotetext[1]{Partially supported by a
  National Key Basic Research Project of China and
  by a grant from NSFC.}

\begin{abstract}
In this paper, we  give a theoretical analysis for the algorithms to
compute functional decomposition for multivariate polynomials based
on differentiation and homogenization which are proposed by Ye, Dai,
Lam (1999) and Faug\`{e}re,  Perret (2006, 2008, 2009).
We show that a degree proper functional decomposition for a set of
randomly decomposable quartic homogenous polynomials can be computed
using the algorithm with high probability. This solves a conjecture
proposed by Ye, Dai, and Lam (1999). We also propose a conjecture
such that the decomposition for a set of polynomials can be computed
from that of its homogenization with high probability. Finally, we
prove that the right decomposition factors for a set of polynomials
can be computed from its right decomposition factor space.
Combining these results together, we prove that the algorithm can
compute a degree proper decomposition for a set of randomly
decomposable quartic polynomials with probability one when the base
field is of characteristic zero, and with probability close to one
when the base field is  a finite field with sufficiently large
number under the assumption that the conjecture is correct.

\end{abstract}

\section{Introduction}
Public key cryptography often relies on a hard mathematical problem.
One of the hard mathematical problems used in cryptosystems is the
functional decomposition problem (FDP) for multivariate polynomials
\cite{2R}.
The general FDP for multivariate polynomials has been proved NP-hard
by Dickerson (1989). Based on this fact, Patarin and Goubin (1997)
proposed $2R$ scheme which is based on the difficulty of decomposing
a set of quartic polynomials.
In the original design, $K$ denotes a finite field of $q$ elements.
The private key consists of:

1. Three linear bijections $r,s,t$: $K^n\rightarrow K^n$.

2. Two quadratic polynomial mappings $\psi,\phi$: $K^n\rightarrow
K^n$.

The public key consists of:

1. The field $K$ and $n$.

2. The composition of polynomial mapping $\pi=t\circ \psi \circ s
\circ \phi\circ r$, which is a set of polynomials of degree four.

In the encryption system, the quadratic polynomials are chosen from
the given S-boxes, which can be inverted easily.  Given the
composition of two quadratic polynomials, if we know the private
key, then we can obtain the plaintext. Otherwise, it is difficult to
invert the polynomials of degree four directly. So, attack on the 2R
scheme is reduced to the functional decomposition of quartic
polynomials.
%

Efficient algorithms for several special forms of FDP  are known.
Polynomial-time algorithms are proposed for univariate decomposition
of multivariate polynomials and multivariate decomposition of
univariate polynomials \cite{muum,von1,von2}.
Efficient algorithms for a kind of monomial decompositions of
rational functions are proposed in \cite{cagd06}, which is further
extended to a complete decomposition algorithm for rational
parametrization of ruled-surfaces \cite{proper-r,proper-r2}.

Ye, Dai, and Lam (1999) proposed an efficient algorithm for
decomposing a set of $n$ polynomials of degree four into two sets of
quadratic polynomials \cite{Dai}. The key idea of computing the FDP
is to differentiate $f$ to obtain a set of cubic polynomials and try
to recover the right decomposition factors from these cubic
polynomials. The idea of differentiation introduced in \cite{Dai} is
a very powerful technique in tackling FDP of multivariate
polynomials.
In a series of papers \cite{Fau06,Fau08,Fau09}, Faug\`{e}re and
Perret made significant contributions to this problem by integrating
the idea of differentiation and fast Gr\"obner basis computation.
In particular, they proposed polynomial-time algorithms for FDP of
semi-regular multivariate polynomial sets. As a consequence, the
current known schemes based on FDP of multivariate polynomials are
considered broken.

As far as we know, the method based on differentiation and
homogenization is the only efficient approach to tackle some of the
general FDP.
But, these algorithms make strong assumptions on the input
polynomial sets and these assumptions are expected to be valid and
can be removed. This paper focuses on the theoretical analysis of
the decomposition algorithm based on differentiation and
homogenization. The main contribution is that  the algorithm can be
used to compute a degree proper decomposition for a set of randomly
 decomposable quartic  homogeneous polynomials with  probability one
when the base field is of characteristic zero, and with probability
close to one when the base field is  a finite field with
sufficiently large number in polynomial time. And if the conjecture
we proposed is correct, it holds for nonhomogeneous case.

We show that a degree proper functional decomposition for a set of
randomly decomposable quartic homogenous polynomials can be computed
using the algorithm with high probability. This solves a conjecture
proposed by Ye, Dai, and Lam (1999). We also propose a conjecture
such that  the decomposition for a set of polynomials can be
computed from that of its homogenization with high probability.
Finally, we prove that the right decomposition factors for a set of
polynomials can be computed from its right decomposition factor
space.
Combining these results together, we prove that if the conjecture is
correct then the algorithm can compute a degree proper decomposition
for a set of randomly decomposable quartic polynomials with
probability one when the base field is of characteristic zero, and
with probability close to one when the base field is  a finite field
with sufficiently large number.

The rest of this paper is organized as follows. In section 2,  we
give the main result.  In sections 3, 4 and 5, we  prove our results
for the three major steps of the algorithm. In section 6, the
algorithm is given and its complexity is analyzed. In section 7, we
conclude the paper by proposing two open problems.

\section{Problem and main result}
In this section, we will present the problem and give the main
results of the paper.

Let $K$ be a field and $\R=K[x_1,\ldots,x_n]$ the polynomial ring in
indeterminates $x_1,\ldots,x_n$ over $K$.
For  natural numbers $u$ and $m$, the functional composition of two
sets of multivariate polynomials
$g=(g_1,\ldots,g_u)$$\in$$K[x_1,\ldots,x_m]^u$ and $ h=(h_1,\ldots,
h_m)$$\in$$\R^m$ is a set of polynomials in $\R^u$:
 \begin{equation}\label{eq-dec1}
 (f_1,\ldots, f_u)=(g_1(h_1,\ldots,h_m),\ldots,g_u(h_1,\ldots, h_m)).
 \end{equation}
  That is,
$$f=g\circ h$$
We call $g$ and $h$ the left and  right {\bf decomposition factors}
of $f$ respectively. The decomposition is called nontrivial if both
$g$ and $h$ contain nonlinear polynomials.

The  {\bf  functional  decomposition problem} (FDP) of multivariate
polynomials is the inverse of the above functional composition
procedure. That is, given a set of $u$ polynomials $f=(f_1,\ldots,
f_u)$$\in$$\R^u$ and a positive number $m$, to find $g=(g_1,\ldots,
g_u)$$\in$ $K[x_1,\ldots,x_m]^u$ and $ h=(h_1,\ldots,
h_m)$$\in$$\R^m $ such that $f=g\circ h$.

It is shown that $f$ always has a nontrivial decomposition when
$m>n$, which is easy to construct \cite{Dickerson}. Then we assume
that $1\leq m\leq n$. Moreover, note that in cryptosystems, the
field $K$ is usually finite and we usually consider the case that
$m=n$. So in the following paper,  assume that $m=n$.

A basic idea of the differentiation  approach is to compute the
linear space generated by the right factors of $f$ from the linear
space generated by certain differentiations of the polynomials in
$f$. For a polynomial sequence $f=(f_1,\ldots,f_u)\in \R^u$ with a
decomposition like \bref{eq-dec1}, let

 $$R_{(f:h)}={\mathrm{span}}_K\{h_1,\ldots,h_n\}$$

be the linear space generated by $h_i$ over $K$, called a {\bf right
factor space} of $f$.

Another idea of the approach is to use homogenization. More precisely,
we first compute a decomposition for the homogenization of $f$
and then try to recover a decomposition of $f$ from this decomposition.
Let $d_f={\mathrm{max}}(d_{f_i})$, $d_g={\mathrm{max}}(d_{g_i})$,
$d_h={\mathrm{max}}(d_{h_i})$.
 The {\bf homogenizations} of $f$, $g$, $h$ are respectively defined
as follows \cite{Fau08,Dai}:
$$f^*=(x_0^{d_f},
 x_0^{d_f}f_1(\frac{x_1}{x_0},\ldots,\frac{x_n}{x_0}),\ldots,x_0^{d_f}f_u(\frac{x_1}{x_0},\ldots,\frac{x_n}{x_0}))  $$
 $$g^*=(x_0^{d_g},
 x_0^{d_g}g_1(\frac{x_1}{x_0},\ldots,\frac{x_n}{x_0}),\ldots,x_0^{d_g}g_u(\frac{x_1}{x_0},\ldots,\frac{x_n}{x_0}))$$
 $$h^*=(x_0^{d_h},
 x_0^{d_h}h_1(\frac{x_1}{x_0},\ldots,\frac{x_n}{x_0}),\ldots,x_0^{d_h}h_n(\frac{x_1}{x_0},\ldots,\frac{x_n}{x_0})).$$

Then the approach proposed in \cite{Fau06,Fau08,Fau09,Dai} can be
divided into three major steps which will be explained later.

\noindent{\bf Algorithm FDPMP}
\begin{itemize}
\item Compute a right factor space $R_{(f^*:h^*)}$ for the homogenization $f^*$ of $f$.

\item Compute a right factor space $R_{(f:h)}$ from $R_{(f^*:h^*)}$.

\item Compute an FDP for $f$ from $R_{(f:h)}$.

\end{itemize}

We will show that there exists a complete polynomial time algorithm
for Step 3, while for Step 1, there exist probabilistic algorithms
in certain cases. We will discuss Steps 1, 2, 3 in the next three
sections.

A decomposition $f=g\circ h$ satisfying the following condition
 \begin{equation}\label{eq-h1} d_f=d_g\cdot d_h\end{equation}
is called a {\bf{degree proper decomposition}},
where $d_f$, $d_g$, and $d_h$ are the degrees of $f$, $g$, and $h$ respectively.
All decompositions in this paper are assumed to be degree proper
unless mentioned otherwise.
In this paper, we will show that the scheme {\bf FDPMP} can be
developed into a polynomial time decomposition algorithm for certain
degree proper decompositions with high probability for random
homogeneous polynomials.  Here, a set of polynomials $f$ is called
{\bf random} or {\bf randomly decomposable} if $f=g\circ h$ and
$g,h$ are random polynomials.

 \begin{theorem}

   Let $f\in K[x_1,\ldots,x_n]^n$ be a set of  quartic homogeneous  polynomials, each polynomial  is of  the same degree,
   for $n\ge 5$, we have a polynomial time probabilistic algorithm
to find a degree proper decomposition $f=g\circ h$ for $g,h\in
K[x_1,\ldots,x_n]^n$. For a random decomposition $f$, the algorithm
will give correct result with probability one when $K$ is of
characteristic zero, and with probability close to one when $K=F_q$
and $q$ is a sufficiently large number.
 \end{theorem}

  If the conjecture proposed in Step 2 is correct, then we have the
  following theorem.

\begin{theorem}\label{th-main}
Let $f\in K[x_1,\ldots,x_n]^n$ be a set of polynomials with degree
less than or equal to four, and at least one polynomial has degree
four. Assume that  Conjecture \ref{p3} is correct, then, for $n\ge
5$, we have a polynomial time probabilistic algorithm to find a
degree proper decomposition $f=g\circ h$ for $g,h\in
K[x_1,\ldots,x_n]^n$. For a random decomposition $f$, the algorithm
will give correct result with probability one when $K$ is of
characteristic zero, and with probability close to one when $K=F_q$
and $q$ is a sufficiently large number.
\end{theorem}

The main idea to prove the above result is to consider the generic FDP.
A {generic polynomial} of degree $d$ in $K[x_1,\ldots,x_n]$ is of
the form $\sum u_{i_{1}\ldots i_{n}}x_{1}^{i_{1}}\cdots
x_{n}^{i_{n}}\,(i_{1}+\cdots+i_{n}\leq d)$ where $u_{i_{1}\ldots
i_{n}}$ are indeterminates.
An FDP $f = g\circ h$ is call a {\bf generic decomposition} if
$g$ and $h$ are generic polynomials of degrees greater than one.

We will show that if $f = g\circ h$ is a generic FDP for two quadratic
polynomials $g$ and $h$, then we can
compute $g$ and $h$ with a polynomial number of arithmetic operations in
the coefficients fields of $g$ and $h$. Furthermore,
when the coefficients of $g$ and $h$ specialize to concrete values
in the base field $K$, the algorithm still works with probability close to one.
%

\begin{remark}\label{rem-1}
Let $N=n(n+1)(n+2)$. Then the coefficients of $g$ and $h$ can be
considered as an element of $K^N$. For convenience, we can also say
that $(g,h)$ is an element of $K^{2N}$.
From the above analysis, if $K$ is of characteristic zero, then the
coefficients of $g$ and $h$ for which the algorithm fails to compute
the decomposition $g\circ h$ consists of an algebraic variety in
$K^N$. In other words, these $(g,h)$ is a subset of $K^N$ with
dimensions lower than $2N$. In this sense, we say that the algorithm
will succeed with probability one.
If $K$ is a finite field, we will give an estimation of the size of
the failure subset and show that it is very small compared  with
$N$.
\end{remark}

\section{Compute an FDP from a right factor space}
\label{sec-fdp} \vskip10pt In this section, we will show how to
compute a decomposition for $f$ from its right factor space
efficiently.
We discuss this problem first, because the result in this section
will be used in Section 5. Also, among the three steps of the
Algorithm {\bf FDPMP}, this is  the only step that has a complete
solution.

We first prove several basic properties for $R_{(f:h)}$.
\begin{lemma}\label{lm-g1}
Two equivalent decompositions of $f$ have the same right factor
space.
\end{lemma}
\emph{Proof}. Suppose that $f$ has two  equivalent decompositions
$g\circ h=g' \circ h'$. By the definition of equivalent
decompositions, there exists  a nonsingular matrix $A\in GL_n(K)$
such that $h'=h\cdot A$. Therefore,
${\mathrm{span}}_K\{h_1,\ldots,h_n\}={\mathrm{span}}_K\{h_1',\ldots,h_n'\}$.\qed

The following result shows that the FDP of a set of polynomials can
be reduced to the FDP  of several single polynomials. Denote the set
of all right factor spaces of $F$ by $SR_F$.

\begin{lemma}\label{lm-g2}
 If $f=(f_1,\ldots,f_u)\in\R^u$, then
 \begin{equation}\label{eq-dec20} SR_f = \cap_{i=1}^u
 SR_{f_i}.\end{equation}
\end{lemma}

\emph{Proof}. It is clear that $SR_f \subseteq \cap_{i=1}^u
 SR_{f_i}$. Assume that $W\in \cap_{i=1}^u
 SR_{f_i}$ and $h_1,\cdots,h_m$ be a basis of $W$. Then there are $g_i\in K[x_1,\cdots,x_m]$
 such that $f_i=g_i(h_1,\cdots,h_m)$. Hence $W\in SR_f$.\qed

Since computing the intersection of two linear spaces is easy, we
may reduce the FDP of $f$ to the FDP of a single polynomial $f_i$.
%


The approaches in \cite{Fau06,Dai} are based on the idea of right
factor space. But, the power of this idea is not fully explained  in
previous work. For instance, it is assumed that the rank of $R_f$ is
$n$ in \cite{Fau08}. It is clear that this condition is not
necessarily correct since $h$ can be a set of arbitrary polynomials.
For instance, if $h=(\sum_{i=1}^n x_i^2,x_2^2,\ldots,x_2^2)$ then
the rank of $R_f$ is always two for any decomposition $f=g\circ h$.

The following result shows that we can recover a right decomposition
factor for $f$ from $R_{(f:h)}$ under any condition.

\begin{theorem}\label{th-basis}
Let $B=\{b_1,\ldots,b_k\}$ be a basis of
$R_{(f:h)}=\mathrm{span}_K$$\{h_1,\ldots,h_n\}$.
If ${\mathrm{dim}}(R_{(f:h)})=k=n$, then  $B$ is a right decomposition
factor of $f$. If ${\mathrm{dim}}(R_{(f:h)})=k<n$, then $
(b_1,\ldots,b_k,b_1,$ $\ldots,b_1)$ is a right decomposition factor of
$f$.
\end{theorem}
\emph{Proof}. Firstly, assume that $\dim(R_{(f:h)})=n$.
Since $\{h_1,\ldots,h_n\}\in R_{(f:h)}$ and $B$ is a basis of $R_{(f:h)}$, each
$h_i$ can be expressed as a linear combination of
$\{b_1,\ldots,b_n\}$. That is, there exists an invertible matrix
$P\in GL_n(K)$ such that $(h_1,\ldots,h_n)=(b_1,\ldots,b_n)\cdot P$.
Then $f=g\circ h=g(X\cdot P)\circ (h\cdot P^{-1})=g(X\cdot P)\circ
(b_1,\ldots,b_n)$, where $X=(x_1,\ldots, x_n)$. Therefore, $B$ is
also a right decomposition factor of $f$.

Secondly, let  $\dim(R_{(f:h)})=k<n$.
For the  decomposition of  $f=g\circ h$, since
$\{h_1,\ldots,h_n\}\in R_{(f:h)}$ and $B$ is a basis of $R_{(f:h)}$,
$h_i=\sum_{j=1}^k a_{i,j}b_j$.

 Therefore, we have
$$(h_1,\ldots,
h_n)=(b_1,\ldots,b _k)\left(
                    \begin{array}{ccc}
                      a_{11} & \ldots & a_{n1} \\
                      \vdots & \ddots & \vdots \\
                     a_{1k} & \ldots & a_{nk} \\
                    \end{array}
                  \right).
$$
and $(a_{ij})_{k\times n}$ contains a  nonsingular $k\times k$ submatrix,
or else $\dim({\mathrm{span}}_K\{h_1,\ldots, h_n\})<k$, a
contradiction.

 Suppose that $$\mathrm{det}\left(
                             \begin{array}{ccc}
                               a_{11} & \ldots & a_{k1} \\
                               \vdots& \ddots & \vdots\\
                               a_{1k} & \ldots& a_{kk} \\
                             \end{array}
                           \right)\neq 0.
$$
Then $(h_1,\ldots, h_n)=(b_1,\ldots,b_k, h_{k+1},\ldots, h_n )A $,
where $$A=                                 \left(
                                                     \begin{array}{cccc}
                                                                      a_{11} & \ldots & a_{k1} &     \\
                                                                      \vdots & \ddots & \vdots &     \\
                                                                      a_{1k} & \ldots& a_{kk} &      \\
                                                                       &  &  &   I_{n-k}   \\
                                                                    \end{array}
                                                                  \right)$$
is an $n\times n$ invertible matrix.
Moreover, let $$B=\left(
         \begin{array}{ccccccc}
            &  &  & -a_{k+1,1}+1 & -a_{k+2,1}+1 & \ldots & -a_{n,1}+1 \\
            &  &  & -a_{k+1,2} & -a_{k+2,2} &\ldots & -a_{n,2} \\
            & I_k &  & \vdots & \vdots & \ddots & \vdots \\
            &  &  & -a_{k+1,k} & -a_{k+2,k} & \ldots & -a_{n,k} \\
            &  &  & &  &  &  \\
            &  &  &  & I_{n-k} &  &  \\
         \end{array}
       \right)
$$
be an $n\times n$ invertible matrix.
It is easy to see that
$$(b_1,\ldots,b_k,b_1,\ldots,b_1)=(b_1,\ldots,b_k, h_{k+1},\ldots,
h_n)B.$$
Hence, $$(h_1,\ldots, h_n)=(b_1,\ldots,b_k,b_1,\ldots,b_1)B^{-1}A.$$
Since $B^{-1}A$ is nonsingular, there exists a $g''$ such that
$f=g''\circ (b_1,\ldots,b_k,b_1,\ldots,b_1)$ which is an equivalent
form of $f=g\circ h$. So we can choose
$(b_1,\ldots,b_k,b_1,\ldots,b_1)$ as a right decomposition factor of
$f$. \qed

Note that  the last $n-k$ elements $b_1$  in the right factor  can be
replaced with any $b_i$ in  Theorem \ref{th-basis}.
\begin{cor}\label{cor-11}
Corresponding to a given right factor space $R_{(f:h)}$, $f$ has a unique
decomposition under the relation of equivalence.
\end{cor}

Restricted to decomposition of quartic polynomials considered in Theorem \ref{th-main},
we have the following result.

\begin{theorem}\label{complexity1}
Use the same assumption as Theorem \ref{th-main}.
If $R_{(f:h)}$ is known, we can compute $g$ with  $O(n^{3\omega})$
arithmetic operations in the field $K$, where $2\le \omega < 3$.
\end{theorem}
{\em Proof.}
Suppose $R_{(f:h)}=\span_K(h_1,\ldots,h_k)$ is known.
Then a right decomposition factor of $f$ is also known by Theorem \ref{th-basis}.
To find $g$, we may simply by solving a system of linear equations
with the coefficients of $g$ as indeterminates. Note that $g$ has
$nC_n^2=O(n^3)$ coefficients. Then we need $O((n^3)^\omega) =
O(n^{3\omega})$ arithmetic operations in $K$ to find $g$, where
$\omega$ is the matrix exponent \cite{lineq} to measure the
complexity of solving linear equations.\qed

\section{Decomposition of a set of homogenous polynomials}
\label{sec-diff}
In this section, we consider the decomposition of $f$ when each
polynomial of it is homogeneous of the same degree.
%
%
More precisely, we will consider the following problem: ``Let $f$ be
a set of quartic  homogeneous polynomials. Find a decomposition
$f=g\circ h$ where $g, h$ are sets of quadratic homogeneous
polynomials."

We may consider the problem in two steps. First, we compute the
following linear space over $K$
\begin{equation}\label{eq-h11}
 \widetilde{V}_f=\span_K\{\frac{\partial f_i}{\partial x_j}: 1\leq i\leq u, 1\leq j\leq n\}.
\end{equation}
Since $f=g\circ h$ and $g$ consists of quadratic polynomials, it is
clear that $\widetilde{V}_f$ is contained in the following linear
space.
\begin{equation}\label{eq-h2}
V_f=\span_K\{x_i h_j:1\leq i, j\leq n\}.
\end{equation}
The following example shows that  $\widetilde{V}_f$ could be a proper subset of
$V_f$.
\begin{example}\label{exm}
Let $f=(xy^2z, x^2y^2+xy^2z, xy^2z+y^2z^2)$, $g=(xz, x^2+xz,
xz+z^2),h=(xy, y^2, yz).$ It is easy to check that $f = g\circ h$.
We have $\widetilde{V}_f=\span_K\{xyz, y^2z, yz^2, xy^2, x^2y\}$ and
$V_f=\span_K\{xyz, y^2z, yz^2, xy^2, x^2y, y^3\}$. $\widetilde{V}_f$  is
a proper subset of $V_f.$ Later in this section, we will see
that $h$ cannot be recovered from its corresponding
$\widetilde{V}_f$ in this example.

\end{example}
%

The idea of the algorithm is to compute $\widetilde{V}_f$ first,
then try to recover $V_f$ from $\widetilde{V}_f$, and finally
compute $R_{(f:h)}$ from $V_f$.
%
%
We will analyze the above procedure in the following two subsections.
The problem is divided into two cases: $u=n$ or $u < n$.

\subsection{The case when $u=n$}

We divide the procedure into two steps: to compute $V_f$ from $\widetilde{V}_f$
and to recover $R_{(f:h)}$ from $V_f$.

{\bf A. Compute $V_f$ from $\widetilde{V}_f$}

When $u=n$,
$\widetilde{V}_f$ is generated by $n^2$ cubic polynomials, and
$\dim(\widetilde{V}_f) \leq \dim(V_f)\leq n^2$.
In the next theorem, we will show that the probability for $\widetilde{V}_f=V_f$
is close to one under some conditions.
The idea of the proof is to find a nonsingular matrix $A$ in some indeterminates
such that if a set of specialization of these indeterminates does not
vanish $|A|$ then $\widetilde{V}_f=V_f$.

\begin{theorem}\label{th-equv}
For randomly chosen $g$ and $h$, let $f=g\circ h$. Then
\begin{enumerate}
   \item
     $\widetilde{V}_f=V_f$ with probability one when the field $K$ is
     of characteristic zero.
   \item
      $\widetilde{V}_f=V_f$ with probability close to one when $K=GF(q)$ and $q$ is sufficiently large.
\end{enumerate}
\end{theorem}
\emph{Proof}. Assume that $$f_i=\sum_{1\leq k, l\leq n}a_{k,l}^{(i)}h_kh_l,
(1\le i \le n) $$ where $a_{k,l}^{(i)}=a_{l,k}^{(i)}$ for $1\leq k,
l\leq n$, and
$$
   h_i=\sum_{1\leq k\leq l\leq
    n}b_{k,l}^{(i)}x_kx_l, (1\le i \le n).
$$
Then
$$
   \frac{\partial f_i}{\partial x_j}=\sum_{1\leq k,
    l\leq n}a_{k,l}^{(i)}(h_k \frac{\partial h_l}{\partial x_j}+h_l
    \frac{\partial h_k}{\partial x_j} ).
$$
Let
$$
  U_i=\left(\frac{\partial f_1}{\partial x_i},\frac{\partial f_2}{\partial x_i},\cdots,\frac{\partial f_n}{\partial x_i}\right),
  V_i=\left(x_ih_1,x_ih_2,\cdots,x_ih_n \right),\,\,\mbox{for}\,\, i=1,\cdots,n.
$$
Let $U=(U_1,U_2,\cdots,U_n)^T$ and $V=(V_1,V_2,\cdots,V_n)^T$.
Each $\frac{\partial
f_i}{\partial x_j}$ can be represented by a linear combination of
$\{x_k h_l, 1\leq k, l\leq n\}$ over $K$ and the coefficients are
expressions in $a^{(i)}_{k,l}, b^{(i)}_{k,l}$. So, there exists an
$n^2\times n^2$ matrix $A$  such that $U=A\cdot V$ where the
elements of $A$ are polynomials in $a^{(i)}_{k,l}, b^{(i)}_{k,l}$.
We will prove the $\det(A)\neq 0$. We make the following substitutions in $A$:
$a^{(i)}_{k,l}=(k+l)^i$ and $b^{(i)}_{k,l}=\delta_{k,l}$ and denote the new matrix by $\bar{A}$, where $\delta_{k,l}$ is
the Kronecker's delta.
After making these substitutions, one has $f_i=\sum_{k,l}(k+l)^ix_k^2x_l^2$ and $h_i=x_i^2$ for $1\leq k,l,i\leq n$.
Now we have
$$
  \frac{\partial f_i}{\partial x_s}=4\sum_{k=1}^n(s+k)^i x_s x_k^2, \,\,\mbox{for}\,\, i,s=1,\cdots,n,
$$
which imply that for all $s=1,\cdots,n$,
\begin{equation*}
\left(
\begin{array}{c}
{\frac{\partial f_1}{\partial x_s}} \\[2mm]
{\frac{\partial f_2}{\partial x_s}}\\[2mm]
\vdots \\[2mm]
{\frac{\partial f_n}{\partial x_s}}

\end{array} \right)=\left(
\begin{array}{cccc}
4(1+s)& 4(2+s)   &  \ldots    & 4(n+s)\\[2mm]
4(1+s)^2&  4(2+s)^2  &  \ldots    & 4(n+s)^2\\[2mm]
 \vdots  &  \vdots &     &     \vdots  \\[2mm]
4(1+s)^n  & 4(2+s)^n & \ldots  & 4(n+s)^n\\

\end{array}
\right)\cdot \left(
\begin{array}{c}
x_sx_1^2\\[2mm]
x_sx_2^2\\[2mm]
\vdots\\[2mm]
x_sx_n^2\\
\end{array}
\right).
\end{equation*}
Therefore $\det(\bar{A})$ is the
products of a constant and $n$ Vandermonde determinants, which is
nonzero. Hence $\det(A)\neq 0$. One can easily see that the total degree of $\det(A)$
in $a^{(i)}_{k,l}, b^{(i)}_{k,l}$ equals $2n ^2$.

When $g$ and $h$ specialize to concrete polynomials in $K[x_1,\ldots,x_n]^n$,
if $A$ is invertible then each element of $V_f$ can be represented
by a linear combination of the elements of $\widetilde{V}_f$. So,
$V_f=\widetilde{V}_f$.

When $K$ is of characteristic zero, $\det(A)\neq 0$ with probability
one in the sense explained in Remark \ref{rem-1}. When $K=GF(q)$,
$\det(A)\neq 0$ with probability at least
$\frac{q-d}{q}=\frac{q-2n^2}{q}$ which is close to one  when $q$ is
sufficiently large \cite{finite field}. These conclude the theorem. \qed

When $\widetilde{V}_f\neq V_f$, Ye et al  proposed a heuristic method to
enlarge $\widetilde{V}_f$, but there is no theoretical guarantee that
the enlarged $\widetilde{V}_f$ is equal to $V_f$ \cite{Dai}.

\vskip  10pt
{\bf B. Recover $R_{(f:h)}$ from $V_f$}

 In this subsection, we assume that  the space $V_f$  is
 known and show how to recover $R_{(f:h)}$ from $V_f$. Given a vector space $V \subseteq K[x_1,\cdots,x_n]$ and
 a set $S\subseteq K[x_1,\cdots,x_n]$, we define $(V:S)=\{h|\forall s \in S,\,\,sh\in V\}.$

By the definition of $V_f$, $x_i h_j\in V_f$ for all $i,j$. Hence,
$h_j\in (V_f:x_i)$, and then $R_{(f:h)}\subseteq (V_f:x_i)$, for all $i$.
So we have $$R_{(f:h)}\subseteq \cap_i(V_f:x_i)=(V_f:L),$$ where $L$ is
the linear space generated by the variables $x_1,\ldots, x_n$.

Note that $R_{(f:h)}\subseteq (\widetilde{V}_f:L)$ does not always hold.
In Example \ref{exm}, $(\widetilde{V}_f:L)=\{yz, xy\}$ while
$R_{(f:h)}=\{yz, xy, y^2\}$. $(\widetilde{V}_f:L)$ is a proper subset of
$R_{(f:h)}$. However, by Theorem \ref{th-equv}, in the general case,
$R_{(f:h)}\subseteq (V_f:L)=(\widetilde{V}_f:L)$ with probability one when
$K$ is of characteristic zero and close to one when $K=GF(q)$ and
$q$ is sufficiently large.

One may ask that
whether $R_{(f:h)}=(V_f:L)$?
It is not always true as shown by the following example.
\begin{example}
Let $f=(x^2y^2, x^4+y^4)$, $g=(xy, x^2+y^2)$ and $h=(x^2,y^2)$.
$(V_f:L)=\span_K\{xy, x^2,y^2\}$. $R_{(f:h)}$ is a proper subset of $(V_f:L)$.
\end{example}
Ye et al proposed a conjecture which  suggests that for random $R_{(f:h)} $,
the two spaces are equal with probability close to one no matter
whether $\dim(R_{(f:h)})=n$ or $\dim(R_{(f:h)})<n$  \cite{Dai}. The
conjecture is as follows:

{\bf {Conjecture Y}}\cite[p319]{Dai}   Let $W$ be a linear space of
dimension $\leq n$ consisting of quadratic forms in $n$ variables
$x_1,\ldots,x_n$, and $L$ be the linear space generated by
$x_1,\ldots,x_n$, $V=\sum_{1\leq i\leq n}x_i W$. For randomly chosen
$W$, the probability $\rho$ that $(V:L)=W$ is very close to one when
$n>2$.

It is one of the theoretical foundations of the differentiation
approach. The authors  \cite{Dai} did not prove it and just gave a
justification with some heuristic arguments.
The  work of Faug\`{e}re and Perret is also based on this basic fact.
When  the number of $(V_f:L)$ equals $n$,  they regarded $(V_f:L)$
as $R_{(f:h)}$ in their algorithm \cite{Fau06,Fau08}.

We will give a proof of the conjecture when $n\geq 5$. Actually, we will extend the conjecture
into a more general case that $W$ and $L$ are linear spaces consisting of homogeneous polynomials with higher degree and
give a proof for this extension of the conjecture.
%
The assumption $n\geq 5$ is not a strict limitation since in
practical usages, $n$ is much larger than five. The number $q$ is
always large in $2R$ or $2R^-$   scheme \cite{Fau06,Fau08}. Before proving
the conjecture, we need a technical lemma. Let $P=(p_1,\cdots,p_n)\in \N^n$. In the following, we will always use $X^P$ to denote
the monomial $x_1^{p_1}\cdots x_n^{p_n}$ and $M(d', x_1,\cdots,x_n)$ to denote the set of all monomials in $x_1,\cdots,x_n$ with degree
$d'$.
%

\begin{lemma}
\label{lem-independent}
Assume that $h_i=\sum_{|P|=d}a_P^{(i)}X^P\in K[a_P^{(i)},x_1,\cdots,x_n]$ are homogeneous polynomials in $x_1,\cdots,x_n$ with degree $d$,
where $i=1,\cdots,n+1$ and $a_P^{(i)}\in K$.  Assume that $d'<d$ and $n\geq 2d$. Then if $\{mh_i | m\in M(d',x_1,\cdots,x_n), i=1,2,\cdots,n+1\}$ are linearly
dependent over $K$, then $\left(a_P^{(i)}\right)$ will vanish a set of polynomials with total degree at most $n{n+d'-1 \choose d'}$.
\end{lemma}
\setlength\parindent{0in} \emph{Proof}. \setlength\parindent{0.24in}
Let us consider $a_P^{(i)}$ as indeterminates for a moment. Assume that $H=\sum c_{m,i}mh_i$ where $c_{m,i}$ are indeterminates.
Regarding $H$ as a polynomial in $x_1,\cdots,x_n$, one can
see that $H$ is a polynomial with $n+d+d'-1 \choose d+d'$ monomials whose coefficients are
polynomials in $c_{m,i},a^{(i)}_P$. Setting $H=0$, one can get a system of the
equations as follows:
$A \vec{c}=0,$
where $A$ is a ${n+d+d'-1\choose d+d'}$ by $n{n+d'-1\choose d'}$ matrix with entries linearly in the $a_P^{(i)}$, and
$\vec{c}=(c_{m_1,1},\cdots,c_{m_j,i},\cdots)$.
By the computation, one can show that ${n+d+d'-1\choose d+d'}>n{n+d'-1\choose d'}$.
Hence $A$ is of full rank if and only if \{$m h_i| m\in M(d',x_1,\cdots,x_n), i=1,\ldots, n+1$\}, are linearly independent. To prove $A$ is of full rank, one only need to prove this for a
specialization of $A$. Since $n\geq 2d$, let $h_1=x_1^d,
h_2=x_2^d,\ldots,h_n=x_n^d,h_{n+1}=x_1x_2\cdots x_d+x_{d+1}x_{d+2}\cdots x_{2d}$.
It leads to a specialization of the matrix $A$. Denote this specialization by $\bar{A}$. We claim that $\bar{A}$ is of full rank, which is equivalent to claim that the polynomials $mx_j^d,
m(x_1\cdots x_d+x_{d+1}\cdots x_{2d}), m\in M(d',x_1,\cdots,x_n),j=1,\cdots,n$, are linearly independent. Assume that
$$\bar{H}=\sum_{m,i}\bar{c}_{m,i}mx_i^d+\sum_m \bar{b}_m m(x_1\cdots x_d+x_{d+1}\cdots x_{2d})=0$$
where $\bar{c}_{m,i},\bar{b}_m\in K$. For convenience, denote $\partial x_1^{q_1} \cdots \partial x_n^{q_n}$ by $\partial m$
where $m=x_1^{q_1}\cdots x_n^{q_n}$. One can see that
\begin{eqnarray*}
 && \frac{\partial^{d+d'} (\bar{H})}{\partial m x_i^d}=
   \begin{cases}
        *\bar{c}_{m,i} & \forall \,\,m'\in M, m'x_1\cdots x_d\neq m x_i^d \,\,\mbox{and} \,\,m'x_{d+1}\cdots x_{2d}\neq m x_i^d ; \\
        *\bar{c}_{m,i}+*\bar{b}_{m'}&  \exists \,\, m'\in M \,\,\mbox{s.t.}\,\, m'x_1\cdots x_d=m x_i^d \,\,\mbox{or} \,\,m'x_{d+1}\cdots x_{2d}=m x_i^d ;
   \end{cases}\\
  && \frac{\partial^{d+d'}(\bar{H})}{\partial m h}=
   \begin{cases}
        *\bar{b}_m &  h=x_1\cdots x_d\,\,\mbox{and}\,\, \forall \,\,m'\in M\,\,\forall \,\ i ,\,\,mx_1\cdots x_d \neq m' x_i^d; \\
        *\bar{c}_{m',i}+*\bar{b}_m &  h=x_{d+1}\cdots x_{2d}\,\,\mbox{and}\,\, \exists\,\, m'\in M\,\,\exists \,\,i \,\,s.t.\,\, mx_1\cdots x_d=m'x_i^d;\\
   \end{cases}
\end{eqnarray*}
where $*$ denote positive integers.
Since $\frac{\partial^{d+d'}(\bar{H})}{\partial \tilde{m}}=0$ for all monomials $\tilde{m}$, the claim is proved. Therefore $A$
is of full rank. Now consider the $a_P^{(i)}$ as the elements in $K$. If $\{mh_i | m\in M(d',x_1,\cdots,x_n), i=1,2,\cdots,n+1\}$ are linearly dependent, which
implies that $A\vec{c}=0$ has a nontrivial solution, then $\left(a_P^{(i)}\right)$ must vanish the determinants of all
$n{n+d'-1\choose d'}$ by $n{n+d'-1\choose d'}$ submatrices of $A$. This completes the proof.  \qed

Let $h=(h_1,h_2,\cdots,h_n)$ where the $h_i$ are homogeneous polynomials with the same degrees in $K[x_1,\cdots,x_n]$ and
let $d_h$ be the degree of $h_i$. Denote $$U(h,d')=\span_K\{mh_i | m\in M(d',x_1,\cdots,x_n), i=1,2,\cdots,n\}.$$
Let $W=\span_K\{h_1,\cdots,h_n\}$. Then we have
\begin{theorem}\label{thm2}
For randomly chosen $h_1,h_2,\cdots,h_n$, if $d'<d_h$ and $n>2d_h$,  then the probability $\rho$ that
$(U(h,d'):x_1^{d'})=W$ is  one when the field $K$ is of characteristic
zero and close to one when $K=GF(q)$ with $q$  sufficiently large.
\end{theorem}

\setlength\parindent{0in}\emph{Proof}. \setlength\parindent{0.24in}
Assume that
$h_i=\sum_{|P|=d_h}a_P^{(i)}X^P\in K[x_1,\cdots,x_n]$, where
the $a_P^{(i)}\in K$. Denote $\overline{U}=\{ H \in U(h,d')|\,\,x_1^{d'}|H\}$.
For $\sum_i{G_i  h_i}\in \overline{U}$, let
$$G_i=\widetilde{G}_{0,i}x_1^{d'}+\widetilde{G}_{1,i}x_1^{d'-1}+\ldots+\widetilde{G}_{d'-2,i}x_1^2+\widetilde{G}_{d'-1,i}x_1+\widetilde{G}_{d',i}$$
and
$$h_i=\widetilde{h}_{0,i}x_1^{d_h}+\widetilde{h}_{1,i}x_1^{d_h-1}+\ldots+\widetilde{h}_{d_h-2,i}x_1^2+\widetilde{h}_{d_h-1,i}x_1+\widetilde{h}_{d_h,i},$$
where $\widetilde{G}_{0,i},\widetilde{G}_{1,i}, \ldots,\widetilde{G}_{d',i}$ are homogeneous
polynomials in $x_2,\ldots,x_n$ with degree $0,1,\ldots,d'$
respectively and $\widetilde{h}_{0,i},\widetilde{h}_{1,i},\ldots, \widetilde{h}_{d_h,i}$
are homogeneous polynomials in $x_2,\ldots,x_n$ with degree
$0,1,\ldots,d_h$ respectively.
Since $\sum_i{G_i  h_i}\equiv 0 \ {\mathrm{mod}}\ x_1^{d'}$,
we have
\begin{eqnarray*}
 \sum_i{G_i  h_i}
 &\equiv &\sum_i
\left(x_1^{d'-1}\left(\widetilde{G}_{1,i}\widetilde{h}_{d_h,i}+\widetilde{G}_{2,i}\widetilde{h}_{d_h-1,i}+\ldots+\widetilde{G}_{d',i}\widetilde{h}_{d_h-d'+1,i}\right)\right.\nonumber\\
&&+x_1^{d'-2}\left(\widetilde{G}_{2,i}\widetilde{h}_{d_h,i}+\widetilde{G}_{3,i}\widetilde{h}_{d_h-1,i}+\ldots+\widetilde{G}_{d',i}\widetilde{h}_{d_h-d'+2,i}\right)+\ldots \nonumber\\
&&+x_1^2\left(\widetilde{G}_{d'-2,i}\widetilde{h}_{d_h,i}+\widetilde{G}_{d'-1,i}
 \widetilde{h}_{d_h-1,i}+\widetilde{G}_{d',i}\widetilde{h}_{d_h-2,i}\right)\\
&& \left.+x_1\left(\widetilde{G}_{d'-1,i}\widetilde{h}_{d_h,i}+\widetilde{G_{d',i}}\widetilde{h}_{d_h-1,i}\right)+\widetilde{G}_{d',i}\widetilde{h}_{d_h,i}\right) \nonumber\\
 &\equiv & 0 \,\,{\mathrm{mod}}\,\, x_1^{d'}. \nonumber
\end{eqnarray*}
Therefore,
\begin{eqnarray}
  \sum_i\left(\widetilde{G}_{1,i}\widetilde{h}_{d_h,i}+\widetilde{G}_{2,i}\widetilde{h}_{d_h-1,i}+\ldots+\widetilde{G}_{d',i}\widetilde{h}_{d_h-d'+1,i}\right) &=& 0, \label{df2}\\
  \sum_i \left(\widetilde{G}_{2,i}\widetilde{h}_{d_h,i}+\widetilde{G}_{3,i}\widetilde{h}_{d_h-1,i}+\ldots+\widetilde{G}_{d',i}\widetilde{h}_{d_h-d'+2,i}\right) & =& 0,\label{df3} \\
  \vdots \nonumber \\
  \sum_i \left(\widetilde{G}_{d'-2,i}\widetilde{h}_{d_h,i}+\widetilde{G}_{d'-1,i}\widetilde{h}_{d_h-1,i}+\widetilde{G}_{d',i}\widetilde{h}_{d_h-2,i}\right)&=& 0,   \label{2} \\
 \sum_i\left(\widetilde{G}_{d'-1,i}\widetilde{h}_{d_h,i}+\widetilde{G_{d',i}}\widetilde{h}_{d_h-1,i}\right)&=& 0,  \label{1}\\
  \sum_i\widetilde{G}_{d',i}\widetilde{h}_{d_h,i} &=& 0.  \label{nan}
\end{eqnarray}
Assume that for each $1\leq k \leq d'$, $\{mx_i^d|m\in M(k,x_2,\cdots,x_n),i=2,\cdots,n\}$ are linearly independent. Then by
the equalities (\ref{df2}) - (\ref{nan}), one has $\widetilde{G}_{j,i}=0$ for $j=1,\cdots,d'$ and $i=1,\cdots,n$. Therefore
$\overline{U}=\{\sum_i\widetilde{G}_{0,i}h_i\}\subseteq W$. Note that $(U(h,d'):x_1^{d'})=\overline{U}$. Hence
$(U(h,d'):x_1^{d'})=W$. By Lemma \ref{lem-independent}, the $a^{(i)}_P$ such that for some $k\leq d'$, $\{mh_i | m\in M(k,x_2,\cdots,x_n), i=1,\cdots,n\}$
are linearly dependent are the zeroes of some polynomials with degree at most $(n-1){n+d+k-2 \choose d+k}\left(\leq (n-1){n+d+d'-2\choose d+d'}\triangleq N\right)$.

Hence when $K$ is of characteristic zero, the probability that
$(U(h,d'):x_1^{d'})=W$ is one;
when $K=GF(q)$, the probability that $(U(h,d'):x_1^{d'})=W$ is at least
$\frac{q-N}{q}$ which is close to one  when $q$ is
sufficiently large \cite{finite field}. \qed

\begin{remark}
   In general, when $K$ is algebraically closed, Theorem \ref{thm2} does not hold for sufficiently large integer $d'$. For randomly chosen $h_1,\cdots,h_n$, the set of zeroes of $\{h_1,\cdots,h_n\}$ in $\P (K)^{n-1}$ is empty, where $\P (K)^{n-1}$ is $n-1$ dimension projective
   space over $K$. Then by the Projective Weak Nullstellensatz Theorem (Theorem 8, p.374, \cite{cox-little-shea}), there is some integer $r$
   such that $<x_1,\cdots,x_n>^r \subseteq <U(h,r-d_h)>$. Let $d'=r-d_h$. Then $M(r,x_1,\cdots,x_n)\subseteq U(h,d')$, which implies that
   $M(d_h,x_1,\cdots,x_n)\subseteq (U(h,d'):x_1^{d'})$. However, in general, $W \neq \span_K(M(d_h,x_1,\cdots,x_n))$.
\end{remark}

\begin{cor}\label{conjecture}
{\bf Conjecture Y} is correct over  $K$ when $n\geq 5$, where $K$ is
of characteristic  zero or is a finite field consisting of a
sufficiently large number of elements.
\end{cor}

As a consequence of Theorem \ref{th-equv} and Corollary \ref{conjecture}, we have
the following result.
\begin{theorem}\label{th-con}
If $f$ is a random decomposition and $n\geq 5$,  then
$(\widetilde{V}_f:L)=R_{(f:h)}$ with probability one when $K$ is of
characteristic zero and with probability close to one when $q$ is
sufficiently large where $K=GF(q)$.
\end{theorem}
Therefore, we can recover $R_{(f:h)}$ from $\widetilde{V}_f$ directly with high probability if
the FDP of $f$ is randomly chosen.

Faug\`{e}re and Perret assumed that $\widetilde{V}_f=V_f$ in their
papers, since they assumed that the decomposition is random, the
dimension of $R_{(f:h)}$ spanned by $h_1,\ldots,h_n$ is $n$, and
$\dim(\widetilde{V}_f)\geq \dim(V_f)$ \cite{Fau06}.

\begin{theorem}\label{complexity2}
Under the same assumptions as Theorem \ref{th-con}.
If $\widetilde{V}_f$ is known, we can compute $R_{(f:h)}$ with
complexity $O(n^{3\omega})$ arithmetic operations in $K$ with
probability one when $K$ is of characteristic zero and with
probability close to one when $q$ is sufficiently large when
$K=GF(q)$.
\end{theorem}
{\em Proof.}  It suffices to randomly choose a linear polynomial $l$
in $x_1,\ldots,x_n$ and compute $(\widetilde{V}_f:l)$ to obtain
$R_{(f:h)}$. Without loss of generality, assume that  $l=x_1+c_2 x_2+\cdots+c_n
x_n$.  Denote it by $X=M_l \cdot Y$.

For all $f\in K[x_1,\ldots,x_n]$, define $M_l(f)=f|_{X=M_l\cdot Y},$
$M_l^{-1}(g)=g|_{Y=M_l^{-1}\cdot X}$, where $g\in
K[y_1,\ldots,y_n]$. Then $M_l^{-1} M_l(f)=f$, and $M_l(f_1
f_2)=M_l(f_1)M_l(f_2)$. So $M_l(l)=l|_{X=M_l\cdot Y}=y_1$. Let
$M_l(\widetilde{V}_f)=\{p|_{X=M_l\cdot Y}:$ for all $ p\in
\widetilde{V}_f\}$.

Then we have $r\in (\widetilde{V}_f:l) \Leftrightarrow rl\in
\widetilde{V}_f\Leftrightarrow M_l(rl)\in
M_l(\widetilde{V}_f)\Leftrightarrow M_l(r)M_l(l)\in
M_l(\widetilde{V}_f)\Leftrightarrow M_l(r)\in
(M_l(\widetilde{V}_f):M_l(l))\Leftrightarrow M_l(r)\in
(M_l(\widetilde{V}_f):y_1) \Leftrightarrow r\in
M_l^{-1}(M_l(\widetilde{V}_f):y_1)$. That is, $
(\widetilde{V}_f:l)=M_l^{-1}(M_l(\widetilde{V}_f):y_1)$.

So in order to compute $(\widetilde{V}_f:l)$, we can first transform
the polynomials in $\widetilde{V}_f$ by a nonsingular coordinate
substitution $X=M_l \cdot Y$ to obtain $M_l(\widetilde{V}_f)$, and
then compute $(M_l(\widetilde{V}_f):y_1)$. Finally, transform
$(M_l(\widetilde{V}_f):y_1)$ to $(\widetilde{V}_f:l)$  by the
inverse transformation $Y=M_l^{-1}\cdot X$. The main arithmetic
complexity relies on the computation of
$(M_l(\widetilde{V}_f):y_1)$.

We construct a matrix $S$ to represent the polynomials of
$M_l(\widetilde{V}_f)$ in a basis of monomials of degree three. Each
row of $S$ corresponds to the coefficients of  each polynomial of
$M_l(\widetilde{V}_f)$  with respect to the monomials of degree
three. Suppose that the monomials are sorted so that the last
$n(n+1)/2$ columns of $S$ correspond to monomials which can be
divided by $y_1$. Then perform linear elimination to $S$, we can
obtain polynomials which can be divided by $y_1$, denoted
 by $t_i, i=1,\ldots,k$, if $n\geq 5$, then $k\leq n$ \cite{Fau09}.  Then $(M_l(\widetilde{V}_f):y_1)=\{t_i/y_1, i=1,\ldots,k\}$. Note that $S$
 is  an $n^2\times C_{n+2}^3$ matrix. Then we need $O((n^3)^\omega) = O(n^{3\omega})$ arithmetic operations to compute $(M_l(\widetilde{V}_f):y_1)$.
The whole  arithmetic complexity of computing $(\widetilde{V}_f:l)$
is also $O(n^{3\omega})$. \qed

\subsection{The case when $u<n$}
We now consider the case of  $u<n$.
In this case,  Faug\`{e}re and
Perret extended $\widetilde{V}_f$ and $V_f$ to new linear spaces
$\widetilde{{V_{fd}}}$ and  ${{V_{fd}}}$:
\begin{eqnarray}
\widetilde{{V_{fd}}}&=&\span_K\{m\frac{\partial f_i}{\partial x_j}:
m\in M(d), 1\leq
i\leq u, 1\leq j\leq n\},\\
V_{fd}&=&\span_K\{m'  h_j:  m'\in M(d+1), 1\leq i, j\leq n\},
\end{eqnarray}
where $M(d)$ represents the set of monomials of degree $d$. It is
obvious that $ \widetilde{{V_{fd}}} \subseteq V_{fd}$. The authors
\cite{Fau06,Fau08} required $\dim(\widetilde{{V_{fd}}})\geq \
\dim(V_{fd})$ by choosing a proper integer $d$,
 which means $ \widetilde{{V_{fd}}} = V_{fd}$.

Assume ${{V_{fd}}}$ is known, and
 try to recover $R_{(f:h)}$ from ${{V_{fd}}}$. By the definition of
${{V_{fd}}}$, $m h_j\in V_{fd}$ for all $m\in M(d+1)$ and $j$.
Hence, $h_j\in (V_{fd}:x_i^{d+1})$, and then $R_{(f:h)}\subseteq
(V_{fd}:x_i^{d+1})$, for all $i$. Hence, $R_{(f:h)}\subseteq \cap_i
(V_{fd}:x_i^{d+1})$. The approach in \cite{Fau06,Fau08,Fau09} makes
use of this property, and recovers $R_{(f:h)}$ by  computing the
quotient $(V_{fd}:x_i^{d+1})$ for some $i$. The authors
\cite{Fau06,Fau08,Fau09} chose that $i=n$.

In the case that $d_g=d_h=2$,  Theorem \ref{thm2} fails.

However, in the general case, if the degrees of $g$ and $h$ are more
than 2, then from Theorem \ref{thm2},  we can compute  $R_{(f:h)}$
by computing the quotient $(V_{fd}:x_i^{d+1})$ when $d+1<d_h$.

From the above discussion, we can see that the results listed above
provide a
theoretical guarantee for the previous work  \cite{
Fau06,Fau08,Fau09,Dai} in certain sense.

\section{Recover the decomposition of $f$ from $f^*$}
\label{sec-homo}

In this section, we  study the relationship between the FDPs of a
set of polynomials $f$ and that of its homogenization $f^*$. We will
show that with high probability, we can recover a decomposition for
$f$ from a decomposition of $f^*$.

For a general FDP $f=g\circ h$, the following result gives the
connection between the FDP of $f$ and the FDP of its homogenization
$f^*$.
\begin{lemma}\label{lm-h0}
  If $f=g\circ h$, then $x_0^{d_gd_h-d_f}f^*=g^*\circ h^*$, where
$d_g, d_h, d_f$ are the degrees of $g$, $h$, $f$ respectively.
\end{lemma}
\emph{Proof}. If $f=g\circ h$, we have $d_g\cdot d_h\geq d_f$.
Hence,
$$f_i(\frac{x_1}{x_0},\ldots,\frac{x_n}{x_0})=g_i(h_1(\frac{x_1}{x_0},\ldots,\frac{x_n}{x_0}),\ldots,h_n(\frac{x_1}{x_0},\ldots,\frac{x_n}{x_0})).$$

Then
\begin{eqnarray*} g^*\circ h^*&=&(x_0^{d_gd_h},
x_0^{d_gd_h}g_1(h_1(\frac{x_1}{x_0},\ldots,\frac{x_n}{x_0}),\ldots,x_0^{d_gd_h}h_n(\frac{x_1}{x_0},\ldots,\frac{x_n}{x_0})),\\
&&
\ldots,x_0^{d_gd_h}g_u(h_1(\frac{x_1}{x_0},\ldots,\frac{x_n}{x_0}),\ldots,h_n(\frac{x_1}{x_0},\ldots,\frac{x_n}{x_0})))\\
&=&(x_0^{d_gd_h},
x_0^{d_gd_h}f_1(\frac{x_1}{x_0},\ldots,\frac{x_n}{x_0}),\ldots,x_0^{d_gd_h}f_u(\frac{x_1}{x_0},\ldots,\frac{x_n}{x_0}))\\
&=&x_0^{d_gd_h-d_f}(x_0^{d_f},
 x_0^{d_f}f_1(\frac{x_1}{x_0},\ldots,\frac{x_n}{x_0}),\ldots,x_0^{d_f}f_u(\frac{x_1}{x_0},\ldots,\frac{x_n}{x_0}))\\
&=&x_0^{d_gd_h-d_f}f^*.\quad\qed
\end{eqnarray*}

As a consequence, we have $(f\circ g)^*=f^*\circ g^*$ if  $d_f \cdot
d_g=d_h$ \cite{Fau08,Dai}.

By  a {\em homogeneous decomposition}  $f=g\circ h$, we mean that
each component of $f$, $g$, and $h$ are homogeneous of the same degree
$d_f$, $d_g$, and $d_h$ respectively.
It is clear that  a homogenous  decomposition is always degree
proper.

The following result gives a necessary and sufficient condition for
$f$ to have an FDP in terms of its homogenization $f^*$.
\begin{theorem}\label{lm-h1}
Let $f=(f_1,\ldots,f_u)\in\R^u$. Then, $f$ has a decomposition if
and only if there exist natural numbers $s$, $t$ such that $x_0^s
f^* = g'\circ h'$ is a  homogeneous decomposition and $x_0^t \in
{\mathrm{span}}_K \{h'_0,\ldots, h'_n\}$.
\end{theorem}
\emph{Proof}. If $f$ has a decomposition $f=g\circ h$, let
$s=x_0^{d_gd_h-d_f}, g'=g^*,h'=h^*, t=d_h$ in Lemma \ref{lm-h0}.
Then the conclusion holds.

We now prove the other direction. If  there are natural numbers $s,t$
such that $x_0^s f^* = g'\circ h'$ is a homogeneous decomposition
and $x_0^t \in {\mathrm{span}_K} \{h'_0,\ldots, h'_n\}$, then
$\deg(h')=t$, $\deg(g')=\frac{s+d_f}{t},$ and we can choose $g'$,
$h'$ such that $x_0^s f^*$ has the following homogeneous
decomposition form by Theorem \ref{th-basis}:
\begin{eqnarray*}
x_0^sf^*&=&(x_0^{s+d_f},x_0^{s+d_f}f_1(\frac{x_1}{x_0},\ldots,\frac{x_n}{x_0}),\ldots,x_0^{s+d_f}f_u(\frac{x_1}{x_0},\ldots,\frac{x_n}{x_0}))\\
&=&(x_0^{\frac{s+d_f}{t}}, g'_1,\ldots, g'_n)\circ (x_0^t,
h'_1,\ldots, h'_n))
\end{eqnarray*}
and $\deg(g'_i)=\frac{s+d_f}{t}$, $\deg(h'_i)=t$.
Let $x_0=1$. We have
\begin{eqnarray*}
f&=&
(f_1,\ldots,f_u)\\
&=& (g'_1(1,x_1,\ldots,x_n),\ldots,g'_u(1,x_1,\ldots,x_n))\circ
(h'_1(1,x_1,\ldots,x_n),\ldots,h'_n(1,x_1,\ldots,x_n)),
\end{eqnarray*}
which is a decomposition of $f$. \qed

As a consequence of Lemma \ref{lm-h0} and Theorem \ref{lm-h1}, we
have

\begin{cor}\label{lm-h2}
Let $f=(f_1,\ldots,f_u)\in\R^u$. Then, $f$ has a degree proper
decomposition if and only if there is a natural number $t$ such that
$f^*$ has a homogeneous decomposition $f^*=g'\circ h'$ and $x_0^t
\in {\mathrm{span}}_K \{h'_0,\ldots, h'_n\}$.
\end{cor}

In order to use the idea of homogenization, we need to solve the
following problem.

\begin{con}\label{p3}
For all homogeneous decompositions of $f^*=G\circ H$, we have
$x_0^{d_H}\in {\mathrm{span}}_K\{H_0,$ $\ldots, H_n\}$.
\end{con}

If the conjecture is true, we may conclude that to compute a degree
proper decomposition of $f$ is equivalent to compute a homogeneous
decomposition of $f^*$. Therefore, we can obtain a right factor
space $R_{(f:h)}$ of $f$ from $R_{(f^*:h^*)}$ in the same way with
the method in the proofs of Theorem \ref{lm-h1}.


\begin{theorem}Conjecture \ref{p3} has a positive answer in the field of complex
numbers if the degrees of $f^*$, $G$ and $H$ are 4,2,2 respectively
and $n=2$.
 \end{theorem}
{\emph{Proof}}. In the field $K=\mathbb{C}$, if $G$ is
nondegenerate, we can assume that $G$ has the following standard
form $G=x_0^2+x_1^2+x_2^2$ by nonsingular linear substitution  (If
$G$ is degenerate, then  we can assume that $G=x_0^2+x_1^2$  or
$G=x_0^2$, it is easy to see that the Conjecture holds in either
case).

Firstly, we claim that we can assume $H_0=x_0^2+c_0$,
$H_1=b_1x_0+c_1$, and $H_2=b_2x_0+c_2$ where $c_i$ are quadratic
homogeneous polynomials and $b_i$ are linear homogeneous polynomials
in variables $x_1$ and $x_2$.
Since we consider the decomposition over the field of complex
numbers, we may assume that $H_k=a_k x_0^2 + G_k (k=0,1,2)$, where
$G_k$ does not contain $x_0^2$. Since $x_0^4=H_0^2+H_1^2+H_2^2$,
$a_0^2+a_1^2+a_2^2=1$. Without loss of generality, we may assume
$a_0^2+a_1^2\ne0$. Let
 $H_0' = \frac{a_1H_1}{\sqrt{a_0^2+a_1^2}} + \frac{a_0H_0}{\sqrt{a_0^2+a_1^2}}$ and
 $H_1' = \frac{a_0H_1}{\sqrt{a_0^2+a_1^2}} - \frac{a_1H_0}{\sqrt{a_0^2+a_1^2}}$. We have
$$H_0^2 + H_1^2 = (H_0')^2 +(H_1')^2$$
and $H_1'$ does not contain the term $x_0^2$. Repeat the above
procedure one more time, we obtain three new polynomials
$H_0'',H_1'',H_2''$ such that $H_1''$ and $H_2''$ do not contain
$x_0^2$.
Since $x_0^4=H_0^2+H_1^2+H_2^2$, we have $H_0''=x_0^2 + b_0x_0+c_0$.
Comparing the coefficients of $x_0^3$, we have $b_0=0$. Thus, the
claim is proved.

Since $x_0^4=H_0^2+H_1^2+H_2^2$,  we have
 $-c_0(c_0+2x_0^2)=H_1^2+H_2^2=(H_1+i H_2)(H_1-i H_2) $. We will
 discuss it in the following  two cases.

 (1) When  $c_0+2x_0^2$  is irreducible, then there exist constants $\alpha, \beta\in K$
 such that $H_1+i H_2=\alpha (c_0+2x_0^2), H_1-i H_2=\beta c_0 $, or $H_1-i H_2=\alpha (c_0+2x_0^2), H_1+i H_2=\beta c_0
 $.
   In either case, we  have $x_0^2 \in \span_K\{H_1, H_2
 \}$.

 (2) When  $c_0+2x_0^2$ is reducible, then there exists a linear
 polynomial $p$ in variables $x_1, x_2$ such that  $c_0+2x_0^2=(\sqrt{2} x_0+p)(\sqrt{2}
 x_0-p)$ where $c_0=-p^2$.

 If $H_1+i H_2$ has a factor $\sqrt{2} x_0+p $ or $\sqrt{2}
 x_0-p$, without loss of generality,  assume $\sqrt{2} x_0+p $ is a factor  of $H_1+i
 H_2$, then there exists   constants $\alpha, \beta\in K$ such that
 $H_1+i H_2=\alpha p (\sqrt{2} x_0+p)$  and $H_1-i H_2=\beta p (\sqrt{2}
 x_0-p)$. Then $p^2\in \span_K\{H_1, H_2\}$. Since
 $H_0=x_0^2+c_0=x_0^2-p^2$, then   $p^2\in \span_K\{H_0,H_1, H_2\}$.

 If   $c_0+2x_0^2$ is a factor of $H_1+i H_2$, then  the same
 as the case (1), $x_0^2 \in \span_K\{H_1, H_2
 \}$.

The above discussion shows that $x_0^2\in {\mathrm{span}}_K\{H_0,$
$H_1, H_2\}$. \qed

 The proof of Conjecture \ref{p3} is still open.
\section{Algorithm and complexity}
Let $f\in K[x_1,\ldots,x_n]^n$ be a set of polynomials with degrees
less than or equal to four, and at least one polynomial has degree
four. We now give the algorithm to find a degree proper
decomposition of $f$. We prove that it is a polynomial time
algorithm with high successful probability if  Conjecture \ref{p3}
is correct. Note that the algorithm is essentially the same as that
given in \cite{Dai}. Our main contribution is the analysis of the
algorithm.

{\bf Algorithm FDPMP4.}

{\bf{Input}}: $f\in K[x_1,\ldots,x_n]^n$ be a set of polynomials
with degrees less than or equal to four, and at least one polynomial
has degree four.

{\bf{Output}}: $g, h\in K[x_1,\ldots,x_n]^n$ such that $f=g\circ h$
is a degree proper decomposition of $f$. The algorithm may fail even
if such a decomposition exists.

 {\bf{Step 1}}. Let
 $f_0^*(x_0,x_1,\ldots,x_n):= x_0^4$,
 $ f_i^*(x_0,x_1,\ldots,x_n):= x_0^4
    f_i(\frac{x_1}{x_0},\ldots,\frac{x_1}{x_0}),   1\leq i\leq
    n,$ and
$\widetilde{V_f}:= \span_K\{\frac{\partial f_i^*}{\partial x_j}:
i,j=0,1,\ldots,n\}$.

{\bf{Step 2}}. Compute $R_{(f:h)}^*:= (\widetilde{V}_f:l)$ as stated in
the proof of Theorem \ref{complexity2}.

%
%
%
%
%
%
%

 {\bf{Step 3}}.
Set $x_0=1$ in $R_{(f:h)}^*$ to obtain  $R_{(f:h)}$:  $R_{(f:h)}:= R_{(f:h)}^*|_{x_0=1}$.

 {\bf{Step 4}}. Perform linear elimination to the generators of $R_{(f:h)}$
 to obtain a basis $(h_1,\ldots,h_k)$ of $R_{(f:h)}$.
If $k=n$, then $h=(h_1,\ldots, h_n)$; otherwise $h=(h_1,\ldots,h_k,
h_1,\ldots,h_1)$.

 {\bf{Step 5}}.
Compute the coefficients of $g$ by solving a system of linear
equations as shown in Theorem \ref{complexity1}.

 \vspace{5mm}

\begin{theorem}
{\bf Algorithm FDPMP4} needs  $O(n^{3\omega})$ arithmetic operations
in the field $K$, where $2\le \omega < 3$. For a random
decomposition $f$, the algorithm computes the decomposition with
probability one when $K$ is of characteristic zero, and with
probability close to one when $K=GF(q)$ $q$ is a sufficiently large
number under the assumption that  Conjecture \ref{p3} is correct.
\end{theorem}

\emph{Proof.} Assume that Conjecture \ref{p3} is correct,  the
complexity of the whole algorithm depends on Step 2 and Step 5, both
of them cost $O(n^{3\omega})$ arithmetic operations by Theorem
\ref{complexity1} and Theorem \ref{complexity2}. Then we have a
polynomial time algorithm to find a degree proper decomposition
$f=g\circ h$ for $g,h\in K[x_1,\ldots,x_n]^n$ with probability  one
when $K$ is of characteristic zero, and with probability close to
one when $K=GF(q)$ $q$ is a sufficiently large number.\qed

This proves  Theorem \ref{th-main}.

\section{Conclusion and problems}
In this paper, we  give a theoretical analysis for the  approaches
of computing functional decomposition for multivariate polynomials
based on differentiation and homogenization proposed in
\cite{Fau06,Fau08,Fau09,Dai}.
We show that a degree proper functional decomposition for a set of
quartic homogenous polynomials can be computed using the algorithm
with high probability from randomly decomposable polynomials. We
proposed a conjecture such that the decomposition for a set of
polynomials can be computed from its homogenization with high
probability. Finally, we prove that the right decomposition factors
for a set of polynomials can be computed from its right
decomposition factor space.
Combining these results together, we show that the algorithm can
compute a degree proper decomposition for a set of quartic randomly
decomposable polynomials with high probability if the conjecture we
proposed is correct.   Conjecture \ref{p3} seems to be correct while
it is unsolved.

Despite of the significant progresses, the general FDP for
multivariate polynomials is widely open. Some of the basic problems
related to FDP of multivariate polynomials are not resolved. We will
give two basic open problems below.

The first problem is about the existence of an algorithm for FDP.

\begin{problem}\label{p1}  Given $f\in \R^n$,
to find an FDP for $f$ is decidable or not.
\end{problem}

Note that in a decomposition $f = g\circ h$, the degrees of $g$ and
$h$ could be arbitrarily high. Consider the following two
transformations:
\begin{eqnarray*}
 T_1&&: (x_1,\ldots,x_n) \Rightarrow (x_1+p,x_2\ldots,x_n)\\
 T_2&&: (x_1,\ldots,x_n) \Rightarrow (x_1-p,x_2\ldots,x_n)
\end{eqnarray*}
where $p$ is a polynomial in $x_2,\ldots,x_n$ of any degree. Then
$T_1\circ T_2=(x_1,\ldots,x_n)$. For any decomposition $f = g\circ
h$, $f = (g\circ T_1) \circ (T_2\circ h)$ is also a decomposition of
$f$. Therefore, one way to solve Problem \ref{p1} is to find the
smallest possible degrees of $g$ and $h$ if a decomposition exists.

The second problem is about the computational complexity of FDP. In
this aspect, even the simplest case is not resolved.

\begin{problem}\label{f2} Let $f\in \R^n$ be a set of quartic
polynomials. Estimate the complexity of computing an FDP of $f$ over
a finite field $K=F_q$. In particular, does there exist a
polynomial-time algorithm for Boolean polynomials?
\end{problem}

{\bf{Remark:}} Faug\`{e}re et al  \cite{Fau10}  also proved the
correctness of  section 4 of our paper, but the corresponding part
of our work was independently finished and used a different method.


\end{document}